\documentstyle[twoside,fleqn,espcrc2]{article}

\newcommand{\AmS}{{\protect\the\textfont2
  A\kern-.1667em\lower.5ex\hbox{M}\kern-.125emS}}

\title{2- and 3-Point Gluon Correlation Functions on the Lattice}

\author{C. 
Parrinello\address{Department of Physics, University of Edinburgh, 
Mayfield Road, Edinburgh EH9 3JZ, UK}} 

\begin{document}

\begin{abstract}
I present some preliminary 
results, obtained in collaboration with C. Bernard and 
A. Soni, for the lattice evaluation of 2- and 3-point gluon correlation 
functions in momentum space, with emphasis on the amputated 3-gluon vertex 
function. The final goal of this approach is the study of the running  
QCD coupling constant as defined from the amputated 3-gluon vertex. 
\end{abstract}
\maketitle
\section{INTRODUCTION}
The task of measuring the QCD coupling constant at energies of the order
of a few GeVs is a major challenge for the lattice community, because
of the deep phenomenological and fundamental implications of such a
measurement.
Many different methods have been proposed to do this, 
 based on the study of the interquark static potential, charmonium spectra
and other quantites \cite{Aida}.
In spite of the success of some of these methods (at least
when applied in the quenched approximation), it would be desirable
to exploit a more fundamental definition of the coupling constant,
arising from the lattice study of some fundamental QCD interaction
vertex, and compare such a direct determination of the coupling with
what is obtained from different methods. We discuss here some  
preliminary results in this direction.
 
In general, the lattice study of gluon (and quark) correlation functions 
provides a valuable tool to gain insight into the non-perturbative QCD 
dynamics at the fundamental level. 
In fact, in this approach one deals with the fundamental degrees of freedom 
of the theory, assuming we are sufficiently close to the continuum limit. 

Using a lattice definition of the gluon, 
 in principle one can evaluate any unrenormalized, 
 complete $n$-point Green's function for the gluon (we omit Lorentz indices):
\begin{equation} 
G_{lat}^{(n)} (x_{1}, \ldots, x_{n}) \ \equiv \ < 
 A_{lat} (x_{1}), \ldots, A_{lat} (x_{n}) > 
\end{equation}
or, in momentum space:
\begin{equation}
G_{lat}^{(n)} (p_{1}, \ldots, p_{n}) \ \delta^{4} (p_{1}+ \ldots +p_{n})
\end{equation}  
Once $G_{lat}^{(2)} (p^2)$ and 
$G_{lat}^{(3)} (p_{1}, p_{2}, p_{3})$ are determined,  
 many interesting issues can be investigated. Firstly, the 
non-perturbative behaviour of the gluon propagator $G^{(2)} (p^2)$, 
which
 has been analyzed by many authors in different approaches
\cite{see}, yet it is still poorly 
understood. 
Secondly, from $G_{lat}^{(2)}$ and $G_{lat}^{(3)}$ one can 
 define non-perturbatively the lattice version of 
the amputated, 1PI 3-gluon vertex function: 
\begin{equation}
\Gamma_{lat}^{(3)} (p_{1}, p_{2}, p_{3}) \equiv 
G_{lat}^{(3)} (p_{1}, p_{2}, p_{3}) 
\times \prod_{i=1}^{3} \ \left[ G_{lat}^{(2)} (p_{i}^2) \right]^{-1} 
\label{eq:ampu}
\end{equation} 
Evaluating the above quantity is the crucial step in order to define the 
QCD coupling constant from the lattice 
3-gluon vertex, as we explain in the following.
 
In order to determine a convenient kinematical setup, 
we consider the general form of the continuum, off-shell 3-gluon vertex 
 \cite{Ball}. Such an expression 
contains 6 independent scalar functions, but for the purpose 
of computing the coupling constant renormalization one only needs to determine 
the function which multiplies the tree-level vertex. Of course this 
is the only one which is divergent when the UV cutoff is removed.

If one evaluates the continuum vertex function $\Gamma_{cont 
\ \alpha \beta \gamma}^{(3)} 
(p_{1}, p_{2}, p_{3})$ 
 at the "asymmetric" point defined by  
\begin{equation}
\alpha = \gamma \not= \beta, \qquad p_{1}=p_{\beta}, \ p_{2}=0, \ p_{3}=-p_{1}
\label{eq:asy}
\end{equation} 
then it can be written as
\begin{equation}
\Gamma_{cont \ \alpha \beta \alpha}^{(3)} (p_{\beta}, 0, - p_{\beta})= -2 \ 
F(p^{2}) \ p_{\beta}
\label{eq:baba}
\end{equation}
The above expression is proportional to the continuum
tree-level vertex evaluated at (\ref{eq:asy}), and the proportionality factor 
 $F(p^{2})$ diverges when removing the UV cutoff. 
One can show that the 
leading term of the 1-loop lattice calculation of the 3-gluon vertex 
for the same kinematics is indeed consistent with (\ref{eq:baba})  
\cite{Hasenfratz}. 
Thus we decide to calculate the lattice vertex function $\Gamma_{lat}^{(3)}$ 
at  
(\ref{eq:asy}) and we set (neglecting terms of higher order in the external 
momentum):
\begin{equation}
\Gamma_{lat \ \alpha \beta \alpha}^{(3)} (p_{\beta}, 0, - p_{\beta})= 
-2 \ F_{lat} (p^2, a) \ p_{\beta}
\end{equation}
where $a$ is the lattice spacing. Then, when $a \rightarrow 0$, 
 one can set 
\begin{equation}
F_{lat} (p^{2}, a) \vert_{p^2 = q^2}= Z_{g}^{-1} (a^2 q^{2}) \ g_{o} (a) 
\end{equation}
for a generic momentum $q^2$. Finally, following \cite{Hasenfratz}, we define 
the renormalized, "running" coupling as 
\begin{equation}
g_{R} (q^{2}) = Z_{A}^{3/2} (a^2 q^{2}) \ Z_{g}^{-1} (a^2 q^{2}) \ g_{o} (a)
\end{equation}
where $Z_{A}$ is obtained from 
the relation
\begin{equation}
G^{(2)}_{lat \ \mu \nu} (p^2) \vert_{p^2 = q^2} = T_{\mu \nu} (q) \ 
Z_{A} (a^2 q^{2})  
 \ \frac{1}{q^2}
\end{equation} 
with $T_{\mu \nu} (q)$ being the projector on transverse fields 
(we will work in the Landau gauge).
\section{THE GLUON PROPAGATOR}
The first step in the above program is the evaluation of  
 $G_{lat}^{(2)} (p^2)$.
 The lattice gluon field can be defined as \cite{Mand}:
\begin{equation}
A_{lat \ \mu} \equiv
\frac{U_{\mu} - U_{\mu}^{\dagger}}{2 i a}
 - \frac{1}{3}
Tr \left(
\frac{U_{\mu} - U_{\mu}^{\dagger}}{2 i a}
  \right)
\label{eq:gluone}
\end{equation}
Earlier lattice studies of $G_{lat}^{(2)} (\vec{p}^2 =0, t) \equiv 
\sum_{i=1}^{3} \ G^{(2)}_{lat \ i i} (\vec{p}^2 =0,t)$ 
\cite{Mand,Gup,Soni} reported  evidence of an effective gluon 
mass which increases with the time
separation, for short time intervals.
More recently, the evaluation of  
 $G_{lat}^{(2)} (p^2) 
\equiv \sum_{\mu=1}^{4} \ 
G^{(2)}_{lat \ \mu \mu} (p^2)$ has allowed a more detailed
investigation of the mechanism of dynamical gluon mass generation and 
other non-perturbative phenomena \cite{us,Testa}. 
Our group \cite{us}
 evaluated $G_{lat}^{(2)} (\vec{p}^2 =0, t)$ and $G_{lat}^{(2)} (p^2)$ 
on different 
sets of (quenched) configurations with $\beta$ ranging 
between 5.7 and 6.3. 

\begin{figure}[htb]
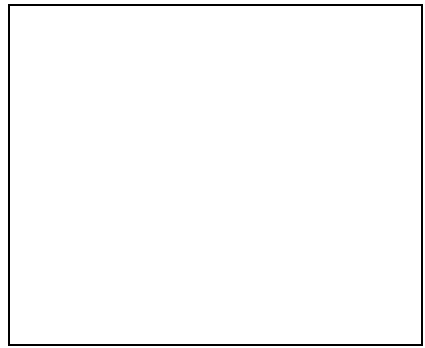

\vspace{9pt}
\framebox[55mm]{\rule[-21mm]{0mm}{43mm}}
\caption{
$G_{lat}^{(2)} (p^2)$ vs. $p$ in GeV on the
$16^3 \times 40$ lattice at $\beta=6.0$. We assume $a^{-1}= 2.1 \ {\rm GeV}$.}
\label{fig:2point}
\end{figure}

For the purpose of defining the amputated vertex function, 
the most convenient set of configurations among the 
ones considered in \cite{us} is the set of 25 configurations on a $16^3 \times 
40$ lattice at $\beta=6.0$. This because it meets requirements of good 
statistics, low infrared cutoff and stable data for the propagator 
(see Fig.\ref{fig:2point}).
All the numerical results shown here refer to such set of configurations.

All the calculations are performed after gauge-fixing to  
 the so-called minimal Landau gauge \cite{Gri,Zwa} (see \cite{us} for a short 
review), 
implemented through
 the iterative minimization of 
\begin{equation} H_U [g] \equiv - {1 \over V} \sum_{n, \mu} \ Re \ Tr \ \left(
U_{\mu}^{g} (n) + U_{\mu}^{g \dagger} (n- \hat{\mu}) \right) 
\label{eq:effecl}
\end{equation}
where $V$ is the lattice volume and $U^{g}$ indicates the gauge-transformed
link
$U_{\mu}^{g} (n) \equiv g (n) U_{\mu} (n) g^{\dagger} (n + \hat{\mu})$. 

\section{THE 3-GLUON VERTEX}
After calculating the propagator, we proceed to the 
evaluation  
of the complete 3-point function $G_{lat \ \alpha \beta \alpha}^{(3)} 
(p_{\beta}, 0, - p_{\beta})$. 
We set $\alpha=1$, $\beta=4$, to be able to inject momentum in the longer 
(time) lattice direction, and we call 
$p_t$ the injected momentum. 
\begin{figure}[htb]
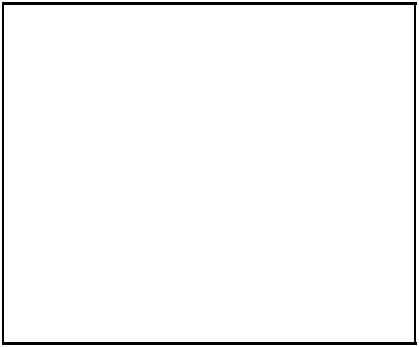

\vspace{9pt}
\framebox[55mm]{\rule[-21mm]{0mm}{43mm}}
\caption{
$G^{(3)}_{lat \ 1 \ 4 \ 1} \ (p_{t}, 0, - p_{t})$ vs. $p_{t}$ in GeV.}
\label{fig:3pointcompl}
\end{figure}
 As expected, the numerical results are consistent with
 an odd function of  
 $p_t$ (see Fig.\ref{fig:3pointcompl}). For large values of $p_t$ 
it gets damped, as a result of the propagators in momentum space 
corresponding to the external legs. 
 At this point, in order to define the amputated 
vertex function, we multiply the complete 
3-point function shown above by the inverse 
propagators, according to (\ref{eq:ampu}).
\begin{figure}[htb]
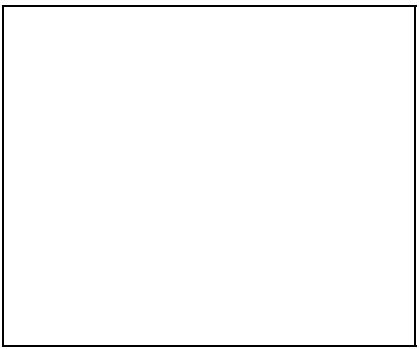

\vspace{9pt}
\framebox[55mm]{\rule[-21mm]{0mm}{43mm}}
\caption{
$\Gamma^{(3)}_{lat \ 1 \ 4 \ 1} \ (p_{t}, 0, -p_{t})$ vs. $p_t$ in Gev.}
\label{fig:amputated}
\end{figure}
 The resulting function $\Gamma^{(3)}_{lat \ 1 \ 4 \ 1}
 \ (p_{t}, 0, -p_{t})$ is shown in 
Fig.\ref{fig:amputated}, where the error is a jackknife one
  The amputated vertex function looks roughly proportional to the external 
momentum, as expected from 1-loop calculations. 
 \section{CONCLUSIONS}
We have discussed a method to measure on the lattice the 3-gluon vertex 
function and to define from it the running coupling constant. 
Preliminary numerical results suggest that  
the vertex function can indeed be non-perturbatively defined 
and measured. A 
 careful analysis of the role of IR and UV lattice 
artifacts is needed. In particular, having defined the coupling at a 
 point in momentum space where one of the external momenta is  
 zero, it will be crucial to investigate finite volume effects. 

C.B. was partially supported by the DOE under grant number
DE2FG02-91ER40628, and C.P. and A.S. were partially supported under USDOE
contract number DE-AC02-76CH00016. C.P. also acknowledges support
from C.N.R. and from SERC under grant GR/J 21347.
 The computing was done at the National
Energy Research Supercomputer Center in part under the
``Grand Challenge'' program and at the San Diego Supercomputer Center.

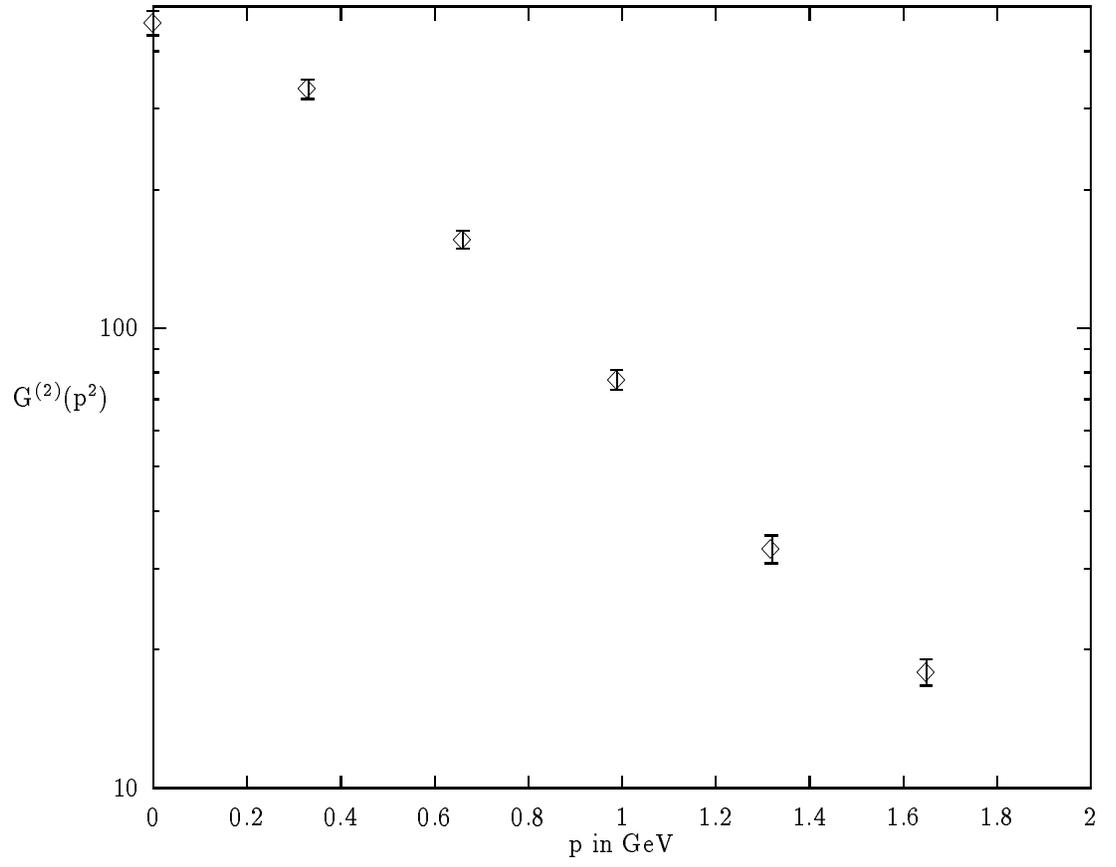
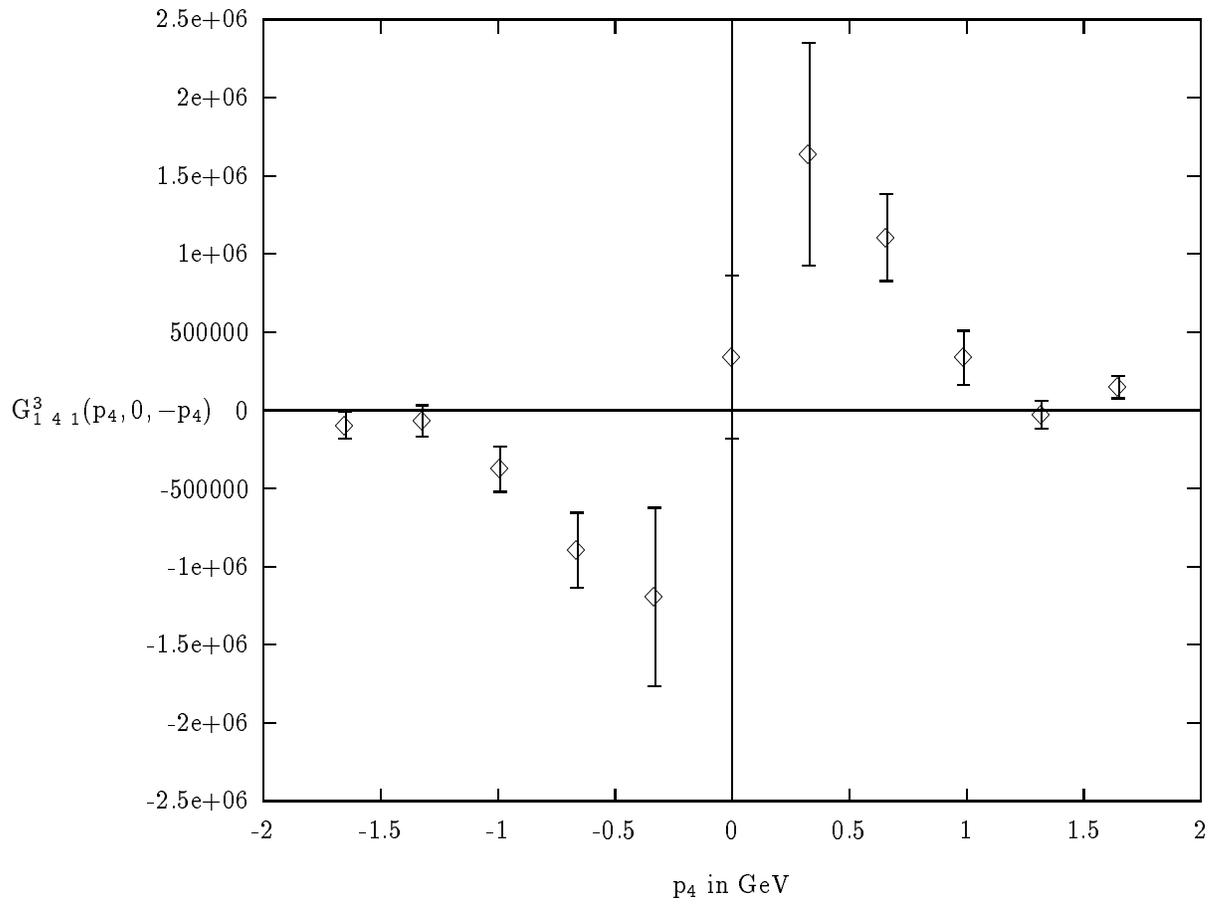
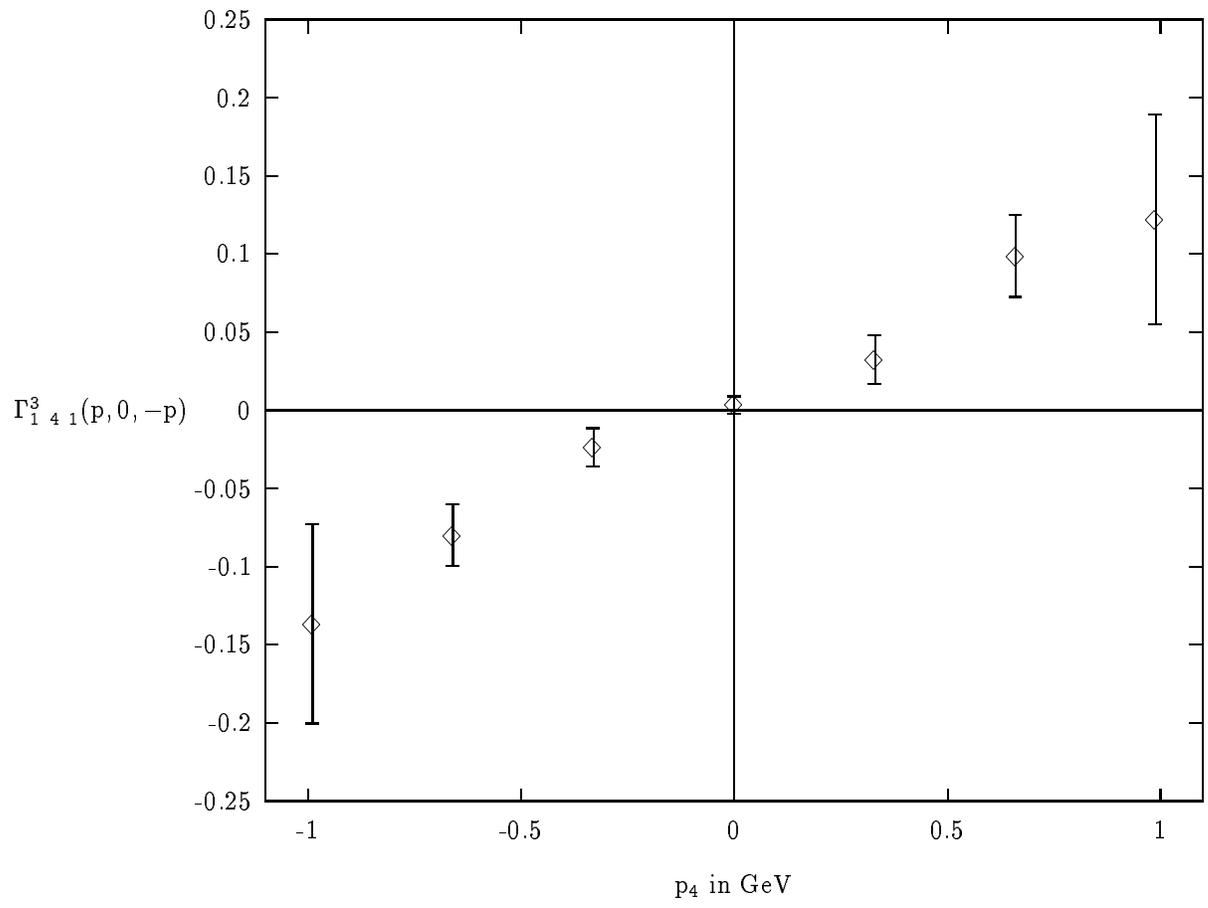
\end{document}